\documentclass[12pt]{article}
\usepackage{amssymb,amsmath}
\usepackage[left=2cm,right=2cm,top=2cm,bottom=2cm]{geometry}
\usepackage{booktabs}
\usepackage{graphicx}
\usepackage{subcaption}
\usepackage{hyperref}
\usepackage{lscape}
\usepackage{enumitem}

\title{Effect of transition layer on wakefield in plasma waveguide}

\author{Peter Markov,\,\,Kostyantyn Galaydych\thanks{Corresponding author: kgalaydych@gmail.com},\,\,Gennadiy Sotnikov}

\date{National Science Center Kharkiv Institute of Physics and Technology 1, Akademichna St., Kharkiv, 61108, Ukraine}

\begin{document}

\maketitle

\section*{Abstract}
Using 2.5D particle-in-cell simulations, we investigate the parameters of a plasma wakefield accelerator (PWFA) for positrons and electrons in a layered plasma waveguide (LPW) in the presence of a transition layer (TL) between two layers of uniform plasma with different densities. The layered plasma is modeled as a combination of a near-wall tubular plasma of higher density and an on-axis plasma column of substantially lower density, separated by a TL in which the plasma density varies smoothly from the lower to the higher value according to a specified law. It is shown that using the PWFA parameters determined for a layered plasma waveguide without the transition layer changes the acceleration conditions when the TL is present, and that this change depends on the TL thickness. It is demonstrated that when the TL thickness exceeds the skin-layer depth in the tubular plasma, the positron acceleration rate decreases substantially. The methods are proposed for correcting the parameters of the plasma waveguide under consideration that compensate for the decrease in the energy gain of the accelerated bunches and ensure their stable transport.

\section{Introduction}\label{sec:1}
Plasma wakefield accelerators are one of the new concepts for accelerating charged particles~\cite{Joshi2019,10.1098/rsta.2019.0215}. In such accelerators, the source exciting the intense field is an electron~\cite{Hogan2016}, positron~\cite{Marsh2003PositronBP}, or proton bunch~\cite{Adli2016}, while plasma is used as the accelerating structure~\cite{Chen1985}. In conventional accelerators, the electromagnetic field is generated in metallic structures by RF sources (klystrons) and is limited in amplitude (tens of MV/m in the GHz frequency range) owing to the development of multipactor discharge and breakdown of the accelerating structures~\cite{Kishek1998,Tazzari2003,Dolgashev2021}. A consequence of this physical limit on the accelerating field is the need for longer linear accelerators to reach higher charged-particle energies. In plasma-based accelerators, the electric field can reach values exceeding by several orders of amplitude the electric field excited in RF-driven accelerators~\cite{osti_4328198,Modena1995}. Thus, research and development into methods for exciting intense wakefields potentially pave the way for creating accelerators that are significantly smaller than conventional ones, for a variety of fundamental and applied studies. The current state of the art and results of PWFA research are presented in the review~\cite{SABERI20261}.

Positron acceleration is one of the current challenges in charged-particle beam physics and accelerator physics~\cite{Cao2024,Si2024,Joshi2025}. In the field of wakefield acceleration, the plasma structures that show the greatest potential specifically for accelerating positron bunches are hollow plasma channels~\cite{Chiou1995,Schroeder1999,Lee2001}. This acceleration method solves the problem of positron defocusing in a uniform plasma when the wakefield is excited by an electron driver~\cite{Blue2003,Corde2015,Cao2024}. In such structures the acceleration rate is high, and the accelerated bunch has a low energy spread at the output~\cite{Chiou1998}; however, these structures are prone to a strong transverse instability caused by the excitation of higher-order wakefield modes in the case of off-axis bunch injection~\cite{Penn2017,Lindstrom2018}. Accordingly, an important research direction for such plasma structures is the development of methods to improve radial stability. Analytical and numerical studies show that one way to achieve transversely stable acceleration is to generate longitudinal and transverse wakefields of comparable amplitude but substantially different (e.g., several-fold different) wavelengths; in dielectric plasma-filled structures the excitation of such fields is weakly coupled (nearly decoupled)~\cite{Sotnikov2014,GALAYDYCH2022,Markov2022,GALAYDYCH2024,SOTNIKOV2025170522}. In~\cite{SotnikovPPCF2026} it was theoretically demonstrated that a two-layer plasma waveguide can potentially provide simultaneous acceleration and focusing of both electron and positron bunches. In that work, the corresponding analytical and numerical studies assumed a sharp boundary between the plasma layers. This approximation makes it possible to derive closed-form analytical expressions for the beam-driven wakefield without using the approximate methods, and to use them to obtain qualitative and quantitative estimates of the amplitudes of its field components. It was found that in such wakefield structure the dominant contribution to the focusing field comes from the bulk Langmuir wave, while the dominant contribution to the accelerating field comes from the eigenmode TM wave of the two-layer plasma waveguide, which is a surface wave. However, a sharp boundary is an idealized model, and under realistic experimental conditions a transition region will exist between plasmas of different densities, i.e., the boundary between them will be diffuse. Since the accelerating mode is of a surface character, the presence of a transition layer can modify both its spatial structure and its amplitude. This, in turn, will affect the dynamics of the driver and accelerated bunch particles. Experimental~\cite{Fan2000,Gessner2016} and numerical~\cite{Bobrova2001,Kimura2011,Bagdasarov2021} studies devoted to creating such plasma structures using laser beams and capillary discharges show that the boundary of a radially bounded plasma is diffuse, and that its transverse density profile can be approximated by smooth model functions to obtain theoretical predictions. Investigating the effect of the transition layer is therefore an important question that was not considered in the original work~\cite{SotnikovPPCF2026}, and it is the main motivation for the present paper.

In practice, the interface is a transition layer whose width is determined by the plasma-formation mechanism (e.g., the laser focusing depth or the ionization diffusion length). This width is typically on the order of tens to hundreds of microns, depending on the specific experimental conditions. If the transition-layer width is much smaller than the plasma skin depth, the actual effect approaches the sharp-boundary assumption. If the transition layer is wider, this can shift the surface-wave frequency, reduce the amplitude of the accelerating field, or introduce additional coupling between the excited modes.

This paper presents the results of simulating the acceleration of positron and electron bunches in the wakefield of an electron driver in a layered plasma waveguide with a transition layer.

\section{Statement of the problem}~\label{sec:2}
\begin{figure}[!tbh]
  \centering
  \includegraphics[width=0.80\textwidth]{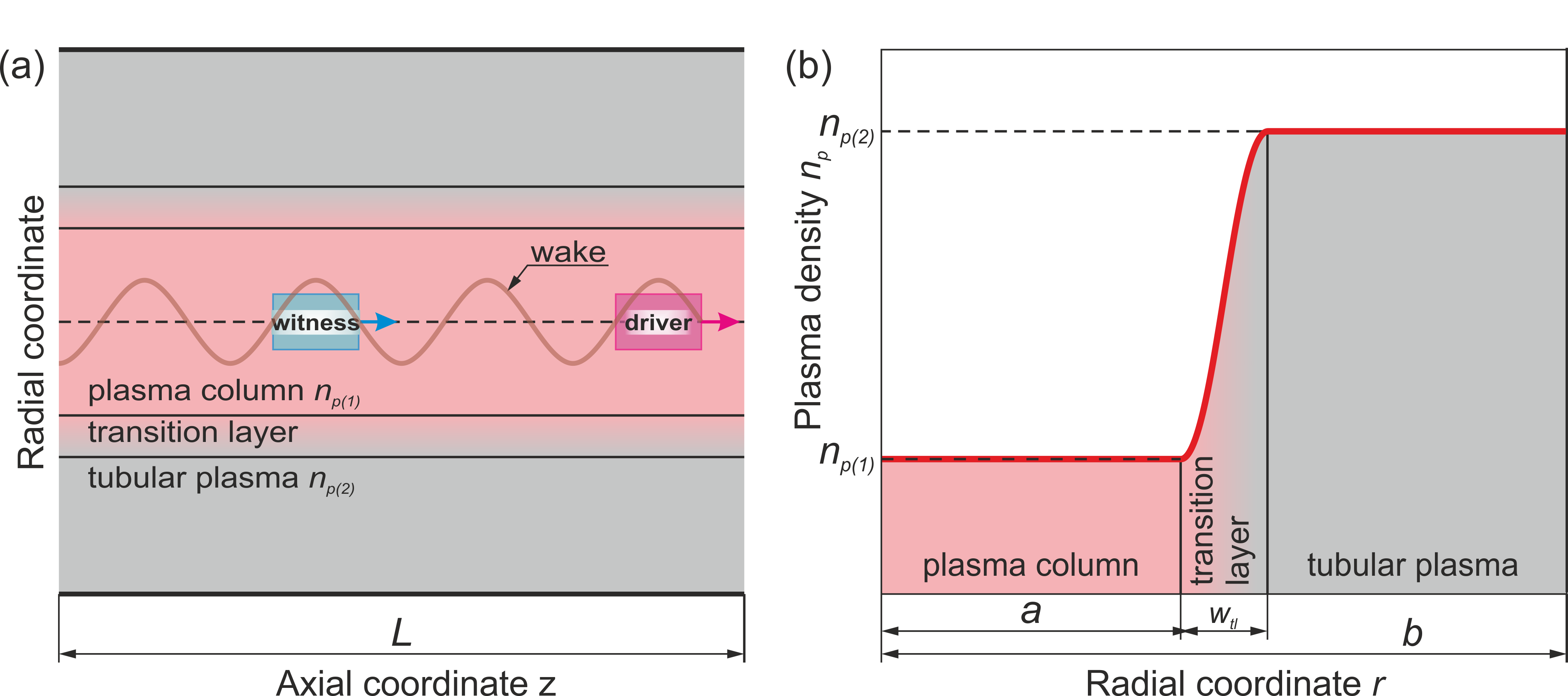}
  \caption{(a) Longitudinal sectional view of cylindrical layered plasma waveguide. Metal coating, layered plasma, transition layer, drive and witness bunches are shown schematically. The drive and witness bunches move along the waveguide axis without an offset. (b) Typical behavior of the plasma density $n_p(r)$ (red curve); $w_{tl}$ is the transition-layer thickness.}\label{Fig:01}
\end{figure}

\begin{figure}[!tbh]
  \centering
  \includegraphics[width=0.40\textwidth]{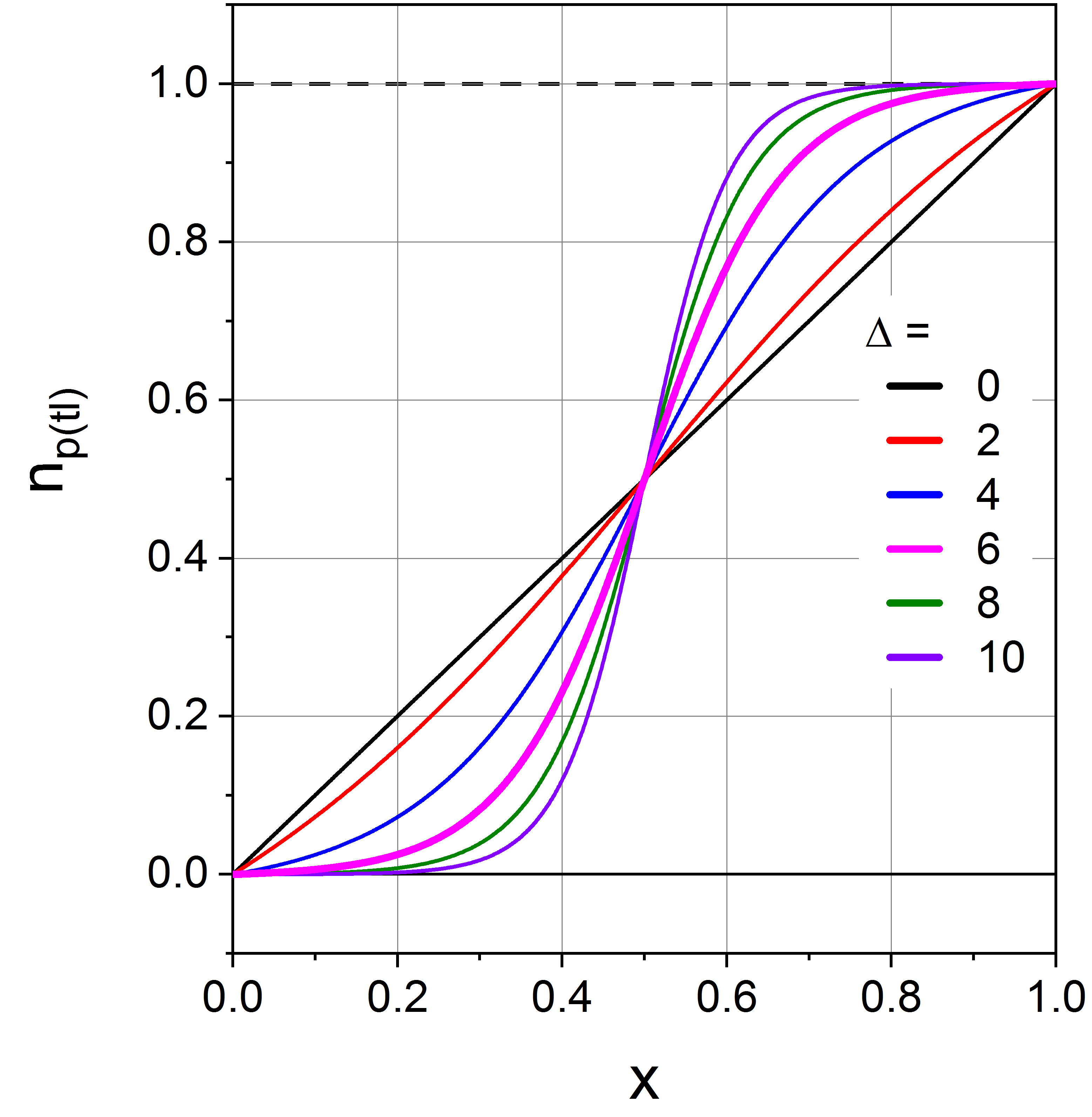}
  \caption{Plot of the function $n_{p(tl)}(x,\,\Delta)$ over the interval $x \in [0,1]$ for several values of the parameter $\Delta$.}\label{Fig:02}
\end{figure}

The wakefield accelerating structure under study is a segment of a cylindrical metallic waveguide of radius $b$ and length $L$, completely filled with a radially inhomogeneous cold plasma (Fig.~\ref{Fig:01}). The radially inhomogeneous plasma consists of three adjoining layers: the inner plasma (a uniform plasma column of radius $a$) has density $n_{p(1)}$, the density of the outer (tubular) uniform plasma is $n_{p(2)}$, and the transition layer with a smooth change in plasma density from $n_{p(1)}$ to $n_{p(2)}$ locates between these two layers. The outer radius of the tubular plasma is equal to the waveguide radius.~\footnote{Filling the waveguide with dense tubular plasma can be realized using laser beams described by high-order Bessel functions~\cite{Fan2000,Kimura2011,Gessner2016}.} No external magnetic field is applied. To excite the wakefield, a drive electron bunch with charge $Q_b$, length $L_b$, and radius $R_b$ is injected into the plasma column. After a certain time delay $t_{del}$ relative to the drive bunch, an accelerated electron/positron witness bunch is injected into the structure. The numerical simulation continues until the drive electron bunch reaches the output of the structure. The aim of the study is to calculate the wakefield excited by the drive electron bunch, to analyze the effect of the transition layer on the wakefield structure, to determine an acceptable relationship between the densities of the plasma layers that allows simultaneous acceleration and focusing of the witness bunches, and to investigate the self-consistent dynamics of the accelerated electron and positron bunches. Although the structure under study differs from the layered plasma waveguide considered previously in~\cite{SotnikovPPCF2026} only by the presence of the transition layer, its parameters may deviate from those of the earlier investigated structure to an increasing extent as the transition layer becomes thicker, in order to achieve the radially stable acceleration regime.
To obtain the required radial dependence of the plasma density within the transition layer, the following smooth function was used in the simulations:
\begin{equation}\label{eq:01}
n_{p(tl)}(x,\,\Delta) =\left( 1- \tanh\frac{\Delta}{2} \right)x + 0.5 \left( \tanh\left[ \left( x-\frac{1}{2} \right)\Delta \right] + \tanh\frac{\Delta}{2} \right),
\end{equation}
where $x$ is the distance from the start of the transition layer, normalized to its width, and $\Delta$ characterizes the deviation of the function $n_{p(tl)}$ from linear behavior. The function $n_{p(tl)}(x,\,\Delta)$ is constructed so that it equals zero at $x=0$, $1/2$ at $x=0.5$, and unity at $x=1$. Figure~\ref{Fig:02} shows the function $n_{p(tl)}$ for $x$ ranging from 0 to 1 for several values of the parameter $\Delta$.

Using the function~(\ref{eq:01}), the analytical expression for the density of the plasma filling the waveguide, as a function of the radial coordinate $r$ and $\Delta$, can then be written as:
\begin{equation}\label{eq:02}
\begin{split}
n_p(r,\,\Delta) =
\begin{cases}
  n_{p(1)},  & \text{$ 0 \le r \le a$},\\
  n_{p(1)} + \left(n_{p(2)} - n_{p(1)}\right)\cdot n_{p(tl)}\left(\frac{r - a}{w_{tl}},\,\Delta\right) ,  & \text{$a < r < a + w_{tl}$},\\
  n_{p(2)},  & \text{$a + w_{tl} \le r \leq b$},
\end{cases}
\end{split}
\end{equation}
where $w_{tl}$ is the transition-layer thickness. In Fig.~\ref{Fig:01}(b), the red curve shows the typical behavior of the dependence $n_p(r)$ given by Eq.~(\ref{eq:02}).

\section{PIC simulation of LPW with transition layer}~\label{sec:3}
For the numerical simulation of wakefield excitation and of the dynamics of the drive electron and accelerated positron and electron bunches, we used our own 2.5-dimensional PIC code SOM~\cite{Markov_2003PJTF,Sotnikov2014,SOTNIKOV2025170522}. The parameters of the LPW (without the transition layer), together with those of the drive electron bunch and the positron/electron witness bunches, are given in Table~\ref{tab:01}. Here the plasma-column density is $4.93\cdot 10^{14}\,\rm cm^{-3}$, which corresponds to a plasma frequency of 199.3~GHz, or a Langmuir wavelength of 1.5~mm. The ratio of the surface (accelerating) wave frequency to the focusing plasma-wave frequency was chosen equal to three, so that the tubular-plasma density is $n_{p(2)}=1.53\cdot 10^{16}\,\rm cm^{-3}$. These parameters are used as the baseline set, against which the optimal parameters obtained for the layered plasma waveguide with the transition layer will be compared.

\begin{table}[!tbh]
\caption{Parameters of the layered plasma waveguide and bunches used in the numerical calculations for the LPW considered in the original paper~\cite{SotnikovPPCF2026}\\[2pt]}
\centering
\begin{tabular}{l l l}
\hline
Parameter & Value &Unit \\
\hline
Waveguide length $L$                                                     & $16$                         & $\rm mm$     \\
Waveguide radius $b$                                                     & $600$                       & $\rm \mu m$     \\
Plasma column radius $a$                                               & $300$                       & $\rm \mu m$     \\
Plasma column density $n_{p(1)}$                                   & $4.93\cdot 10^{14}$   & $\rm cm^{-3}$   \\
Tubular plasma density $n_{p(2)}$                                   & $1.53\cdot 10^{16}$   & $\rm cm^{-3}$   \\
Energy of bunches                                                          & $5$                           & $\rm GeV$       \\
Radius of bunches $R_b$                                                & $250$                       & $\rm \mu m$     \\
Charge of drive electron bunch $Q_b$                               & $-2$                         & $\rm nC$        \\
Charge of witness positron/electron bunch                         & $\pm0.2$                  & $\rm nC$        \\
Length of drive bunch                                                        & $250$                       & $\rm \mu m$     \\
Length of witness bunch                                                    & $125$                       & $\rm \mu m$     \\
Longitudinal rms deviation of drive electron bunch charge     & $125.33$                  & $\rm \mu m$\\
Longitudinal rms deviation of witness bunch charges            & $62.66$                    & $\rm \mu m$\\
Positron witness bunch injection delay time $t_{del}^{e^+}$  & $3.63$                      & $\rm ps$   \\
Electron witness bunch injection delay time $t_{del_{\mathstrut}}^{e^-}$ & $1.17$    & $\rm ps$   \\
\hline
\end{tabular}
\label{tab:01}
\end{table}

For the numerical analysis, three different transition-layer thicknesses were considered: $w_{tl}^{(1)} = 20\,\rm \mu m$, $w_{tl}^{(2)} = 50\,\rm \mu m$, and $w_{tl}^{(3)} = 100\,\rm \mu m$, with a nonlinear variation of the plasma density with radius $r$ (the parameter $\Delta$ in Eq.~(\ref{eq:01}) was set to $\Delta = 6$, see bold magenta curve in Fig.~\ref{Fig:02}). The waveguide radius $b$ was kept unchanged relative to the baseline case, while the radii of the plasma column and of the tubular plasma were each reduced by $w_{tl}/2$.

\begin{figure}[!tbh]
  \centering
  \includegraphics[width=0.90\textwidth]{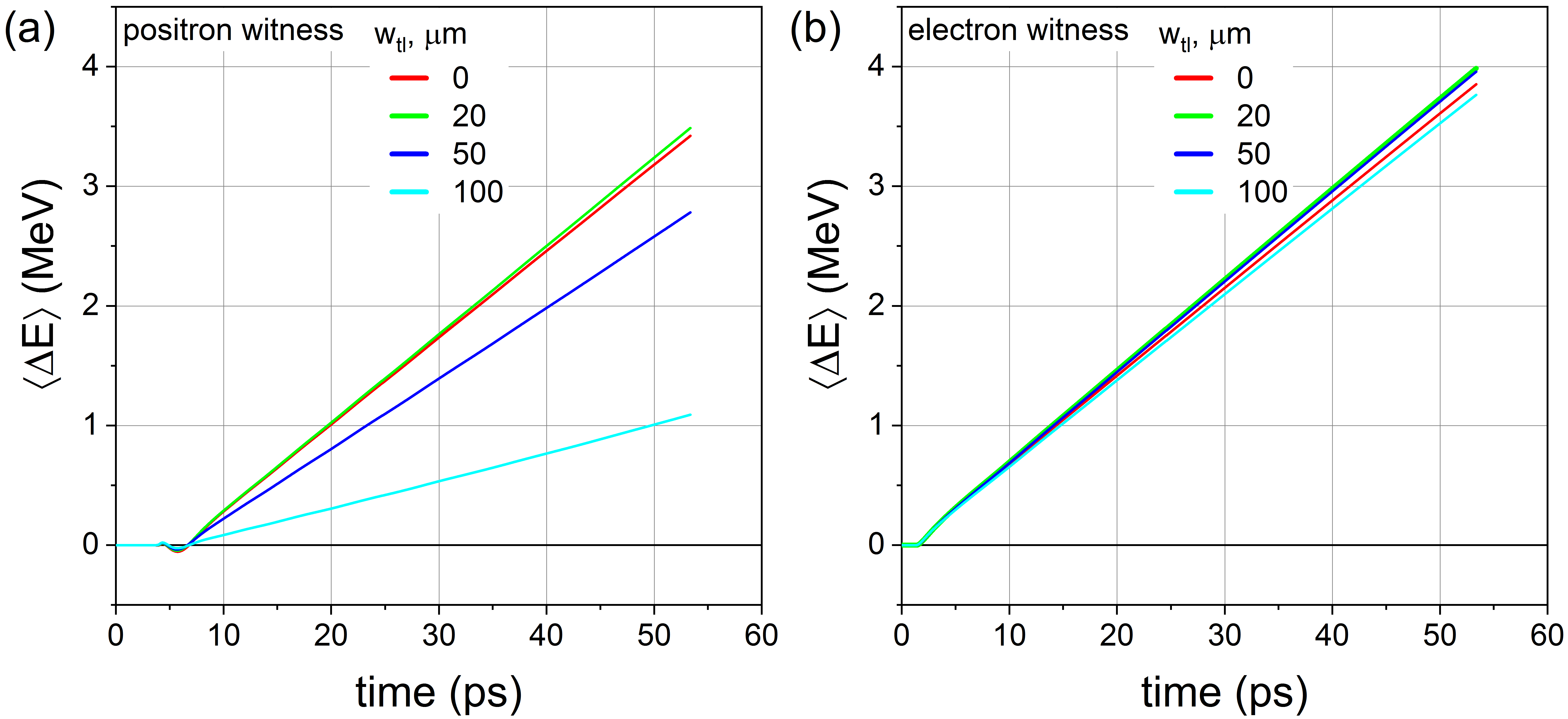}
  \caption{Change in the mean energy of the macroparticles of the positron (a) and electron (b) witness bunches as they move in the wakefield generated by the electron driver, for different transition-layer widths.}\label{Fig:03}
\end{figure}

Figure~\ref{Fig:03} shows the change in the mean energy gain $\langle\Delta E\rangle$~\footnote{The change in the mean energy of the bunch macroparticles was calculated as $\langle\Delta E\rangle = \sum_{i=1}^{N}q_i\Delta E_i/Q$, where $\Delta E_i = \gamma_i - \gamma_{0i}$, $q_i$ is the charge, $\gamma_{0i}$ and $\gamma_i$ are the initial and current relativistic factors of the macroparticle, $Q$ is the total bunch charge, and $N$ is the number of macroparticles in the bunch.} of the positron (a) and electron (b) witness bunches as a function of time as they move in the wakefield generated by the electron driver, for different transition-layer widths, with all simulation parameters except the plasma-column radius $a$ kept the same as in Table~\ref{tab:01}. Before time $t_{del}$, no accelerated bunches are present in the structure. After the accelerated bunches are injected, a quasi-static state is established over some time, during which the bunch first passes through the accelerating phase and then through the decelerating phase of the axial field $E_z$. Once the quasi-static state has formed, the witness bunch move into the regular accelerating phase of the axial field $E_z$, and the energy gain becomes an almost linear function of time. The evolution of $\langle\Delta E\rangle$ shown in the figure continues until the drive bunch reaches the output of the structure (at $t = 53.37$~ns), at which point the simulation is stopped. The case $w_{tl} = 0$, shown for comparison, corresponds to a sharp change in plasma density at $r = a$, i.e., without the transition layer. It can be seen that for $w_{tl}^{(1)} = 20\,\rm \mu m$ the result differs only slightly from the sharp-boundary case. Increasing the transition-layer thickness to $w_{tl}^{(2)} = 50\,\rm \mu m$ reduces the mean energy gain of the positron witness particles by a factor of 1.23. The widest transition layer considered, $w_{tl}^{(3)} = 100\,\rm \mu m$, reduces the positron witness acceleration even further, by a factor of 3.14. For the same transition-layer widths, the corresponding changes in the energy gain of the accelerated electron bunch are negligible (about 100 times smaller).

The reason for the substantial degradation of positron witness acceleration with increasing TL thickness can be explained as follows: the plasma density within the TL is lower than $n_{p(2)}$. As a result, the resonant frequency of the accelerating wave decreases as the TL thickness increases. The regularity of the oscillations responsible for focusing also decreases slightly. The reduction in the axial electric field frequency causes a desynchronization between the acceleration phase of the positron witness bunch and the phase in which it is focused by the transverse wakefield. Consequently, for a fixed delay time $t_{del}^{e^+}$ of the positron witness bunch (as given in Table~\ref{tab:01}), its acceleration decreases, and for a large TL thickness the acceleration can even turn into deceleration. The absence of such a strong dependence between the electron witness acceleration and the TL thickness is related to the short injection delay time $t_{del}^{e^-}$ of the electron witness bunch, which is substantially smaller than $t_{del}^{e^+}$. As a result, the desynchronization between the acceleration phase of the electron witness bunch in the axial field and its focusing phase in the radial field is weak, so the reduction in its acceleration is correspondingly small.

\begin{figure}[!tbh]
  \centering
  \includegraphics[width=0.24\textwidth]{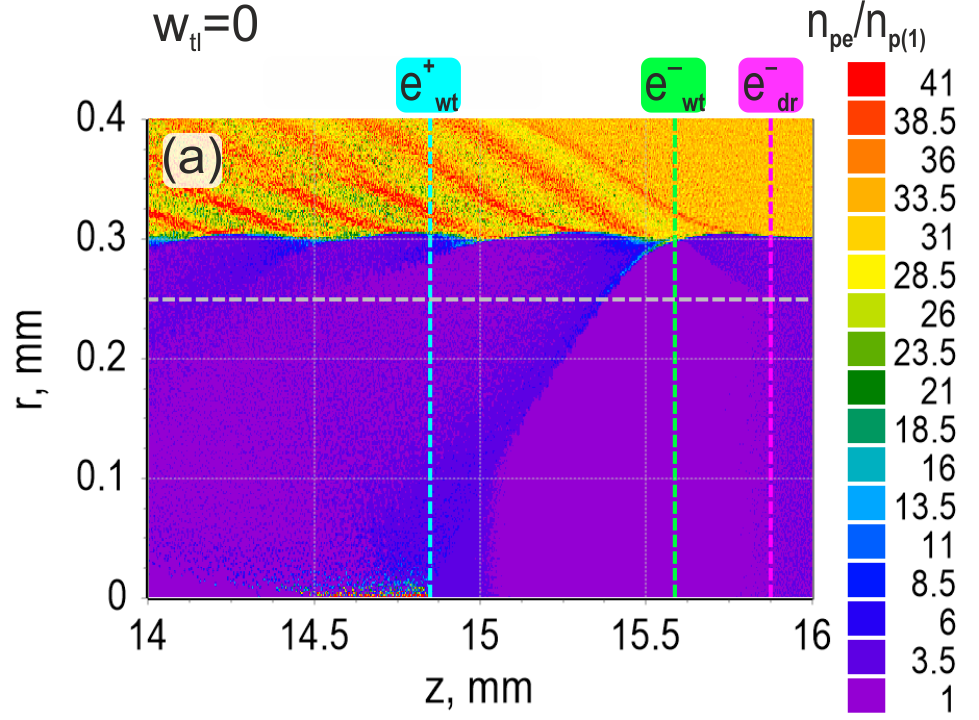}
  \includegraphics[width=0.24\textwidth]{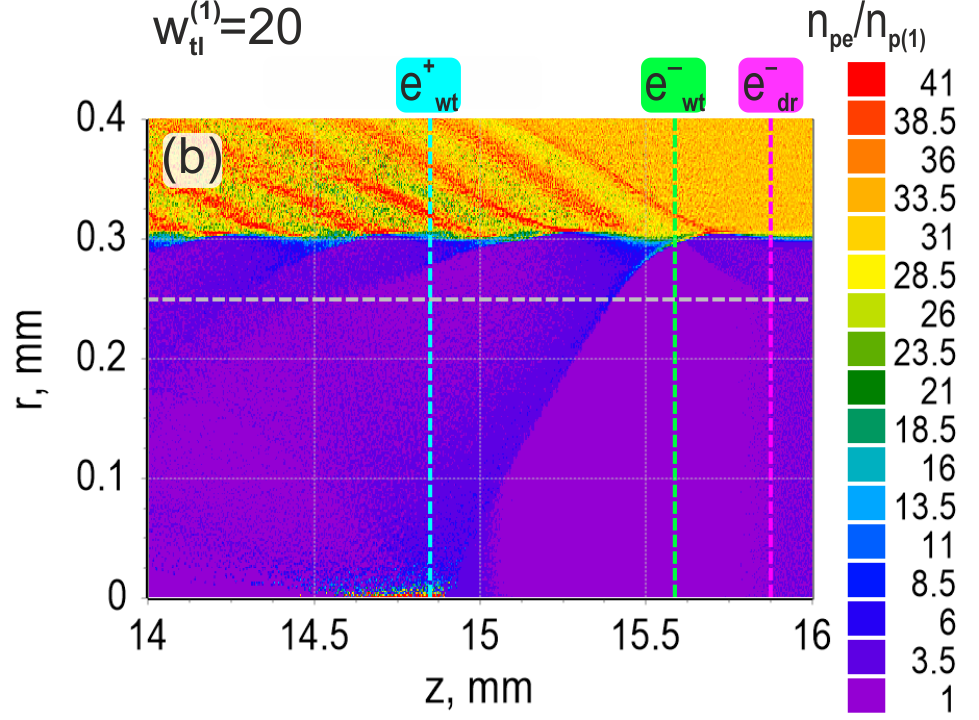}
  \includegraphics[width=0.24\textwidth]{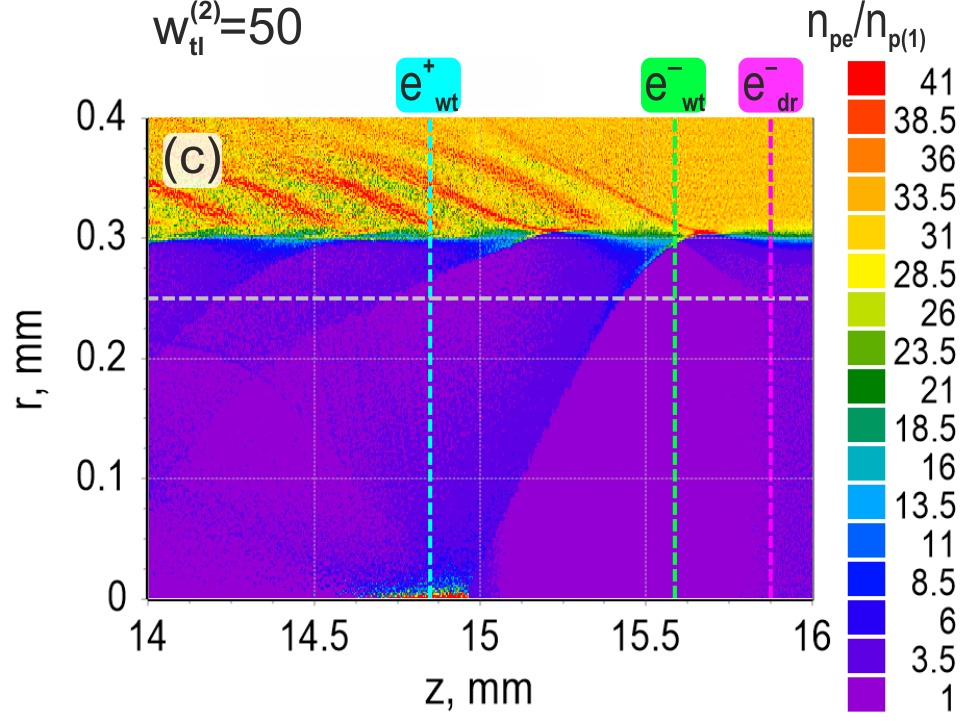}
  \includegraphics[width=0.24\textwidth]{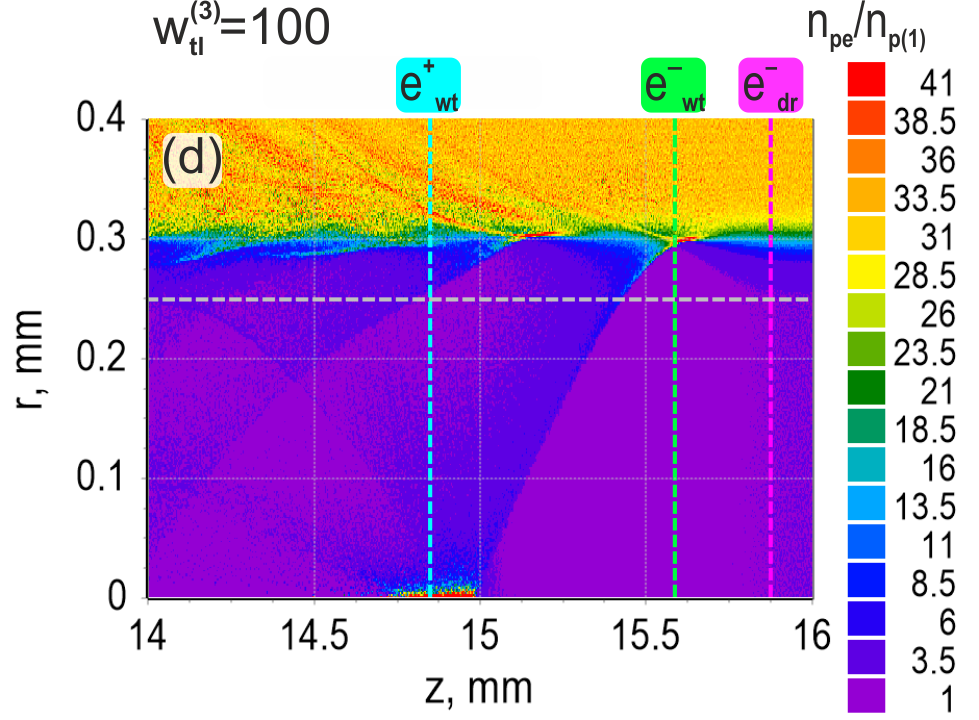}\\
  \includegraphics[width=0.24\textwidth]{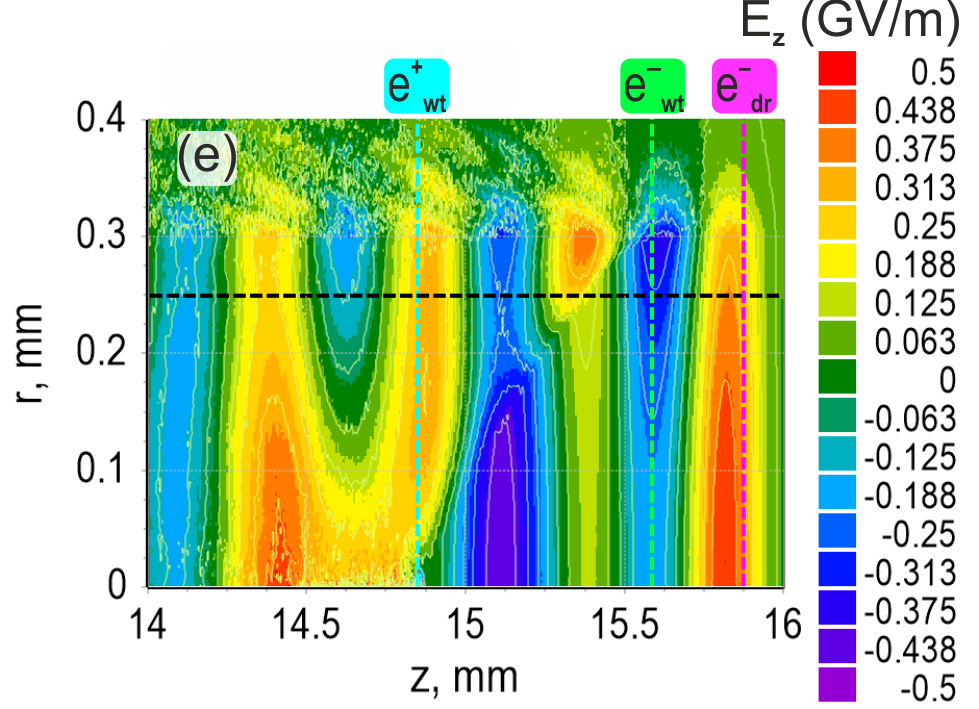}
  \includegraphics[width=0.24\textwidth]{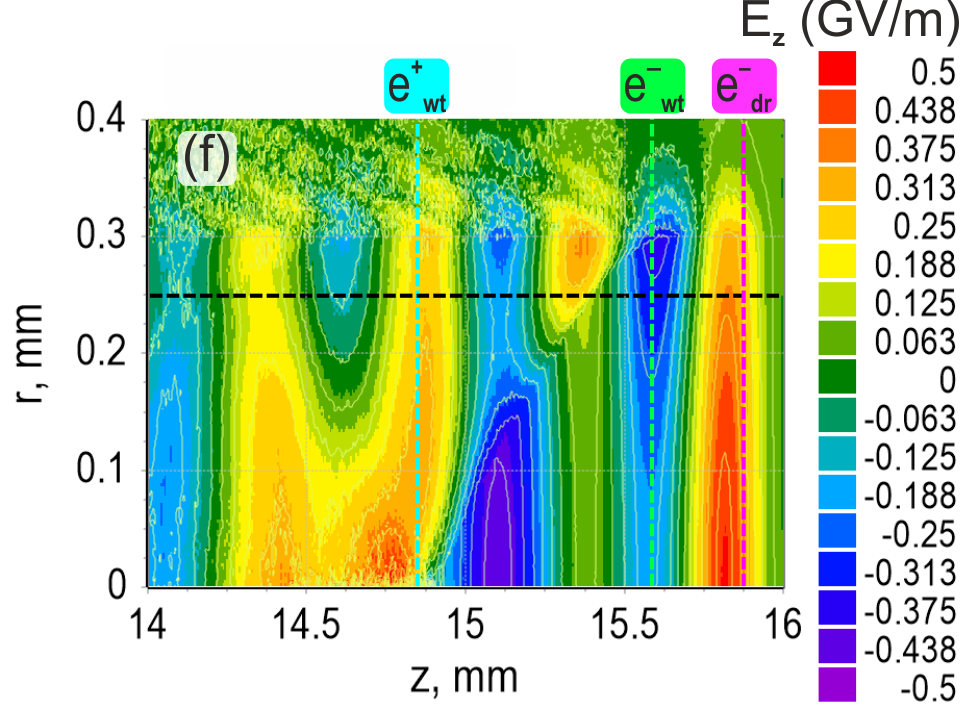}
  \includegraphics[width=0.24\textwidth]{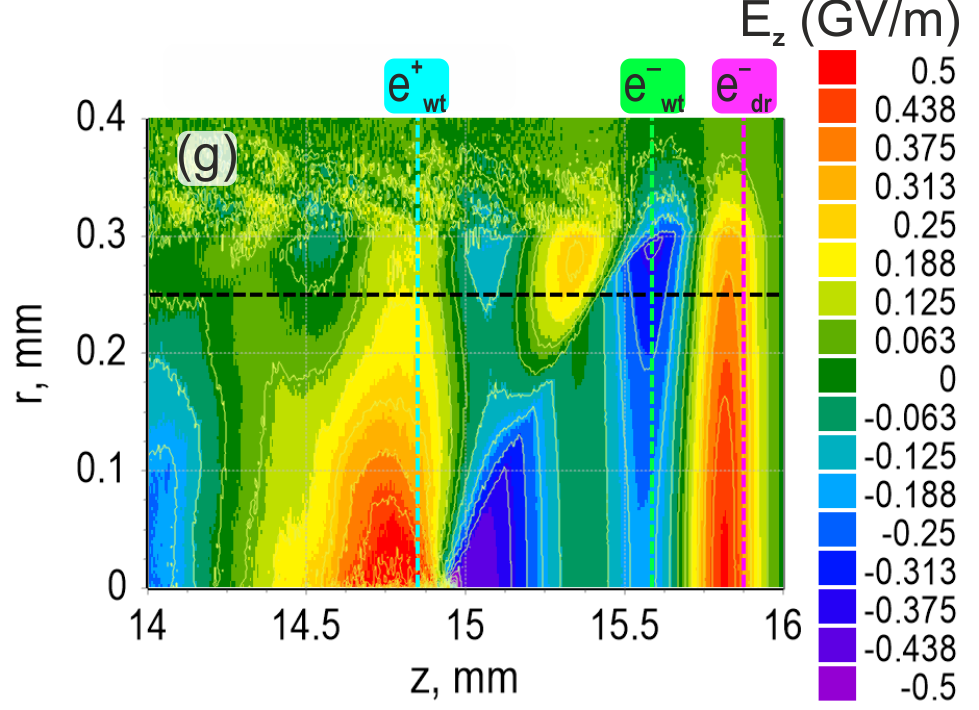}
  \includegraphics[width=0.24\textwidth]{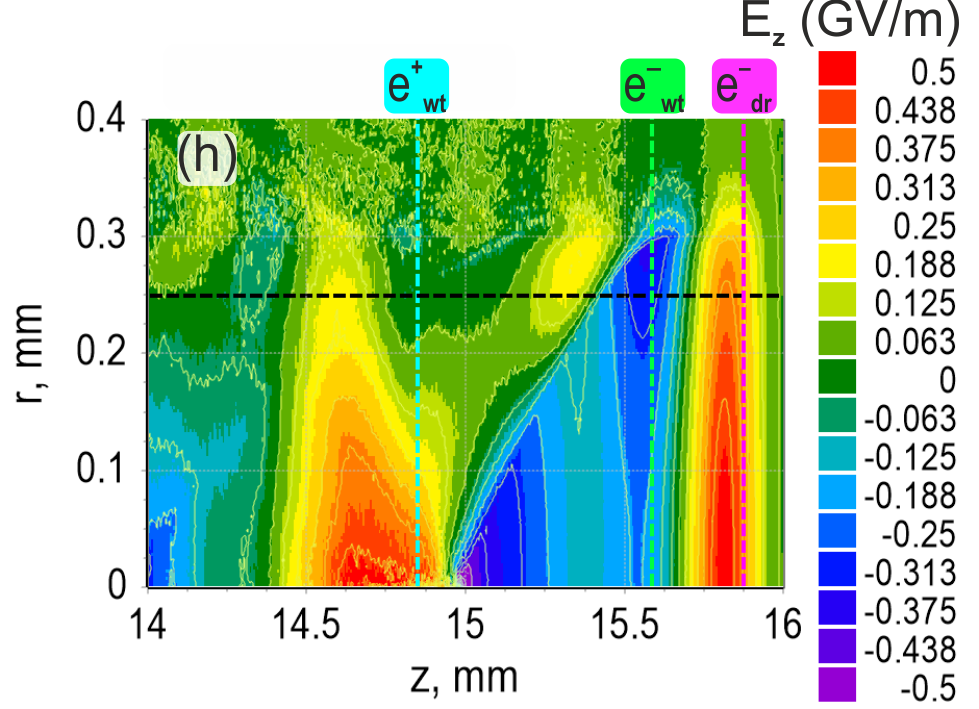}\\
  \includegraphics[width=0.24\textwidth]{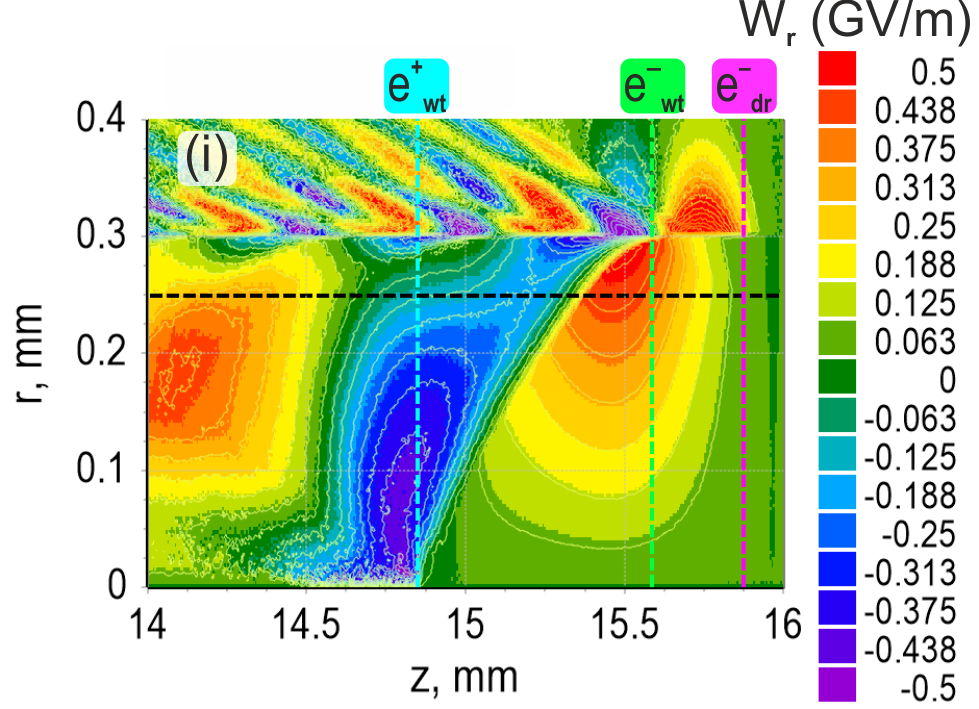}
  \includegraphics[width=0.24\textwidth]{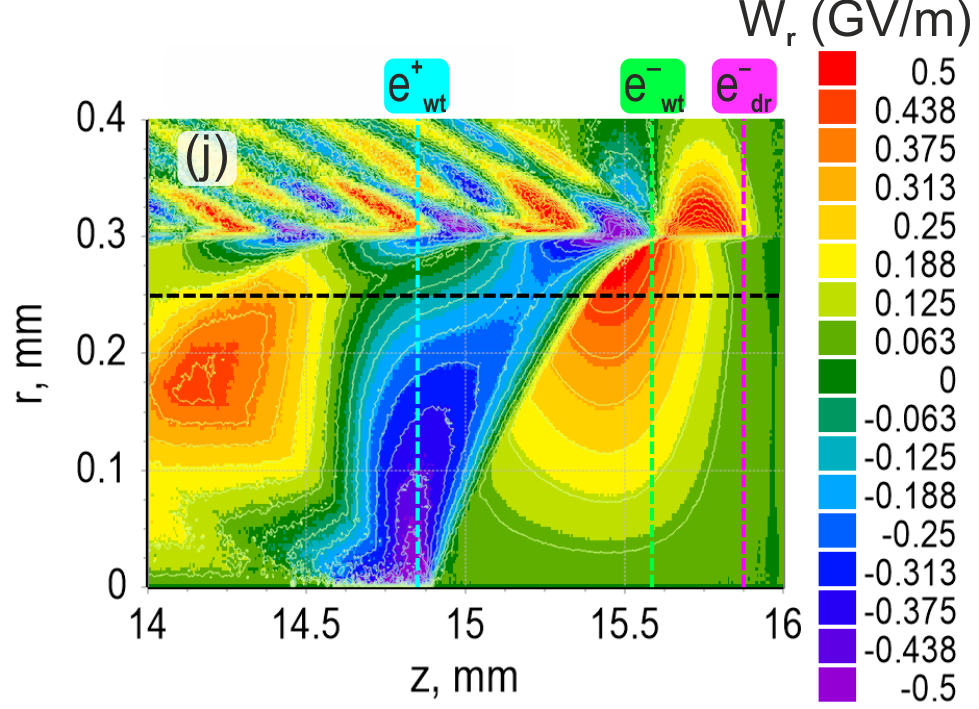}
  \includegraphics[width=0.24\textwidth]{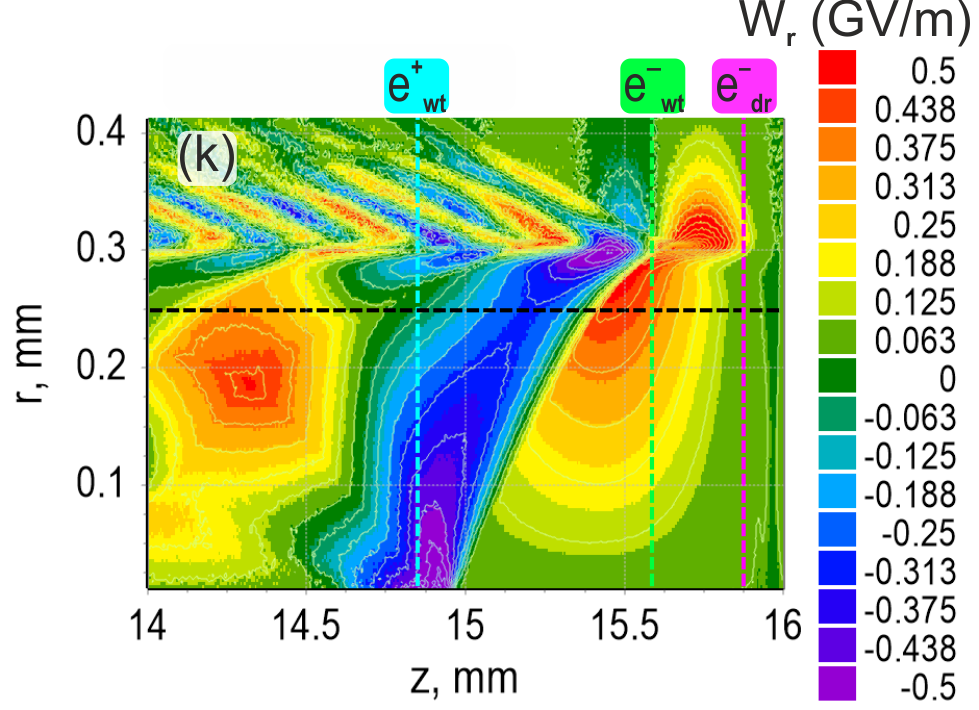}
  \includegraphics[width=0.24\textwidth]{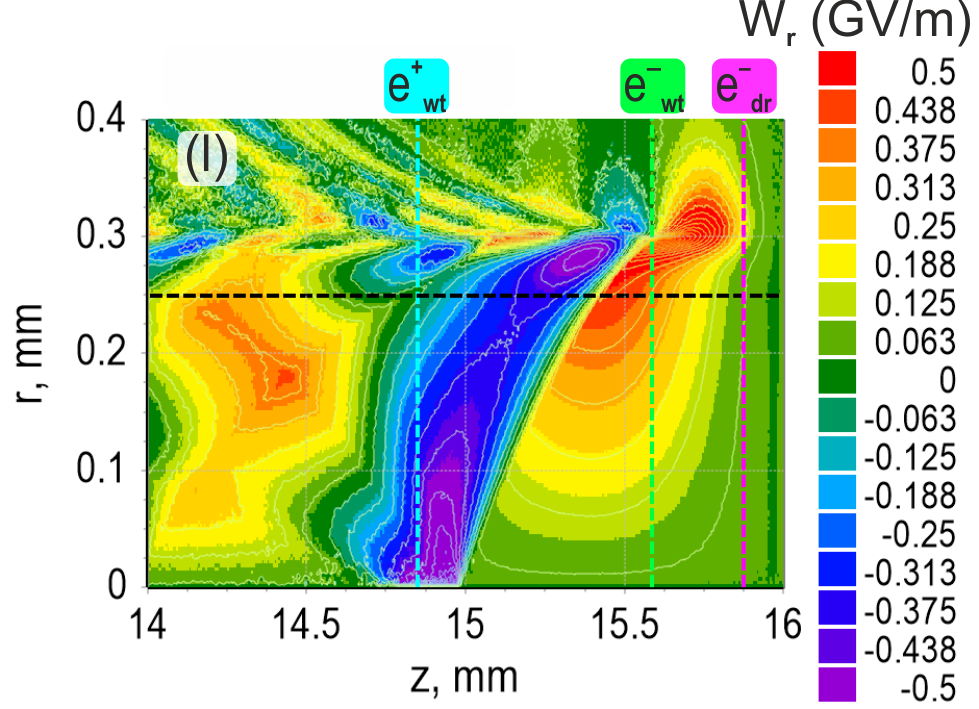}\\
  \caption{Color maps of the plasma electron density $n_{pe}/n_{p(1)}$ and of the axial ($E_z$) and radial ($W_r$) wakefields excited by the electron driver, as functions of the longitudinal ($z$) and transverse ($r$) coordinates. Vertical dashed lines with their captions mark the middle positions of the witness bunches (cyan color for positrons and green color for electrons) and driver bunches (magenta). The horizontal dashed line shows the bunch radius. The columns correspond to different transition-layer widths ($w_{tl}$).}\label{Fig:04}
\end{figure}
These arguments are confirmed by Fig.~\ref{Fig:04}, which shows color maps of the plasma electron density $n_{pe}/n_{p(1)}$ (Fig.~\ref{Fig:04}(a), (b), (c) and (d)), the axial electric field $E_z(r,\,z)$ (Fig.~\ref{Fig:04}(e), (f), (g) and (h)), and the radial wakefield $W_r(r,\,z) = E_r - \beta H_\varphi$ (Fig.~\ref{Fig:04}(i), (j), (k) and (l)) excited by the drive bunch. The vertical dashed lines with their labels mark the mean positions of the witness bunches (cyan for the positron bunch, green for the electron bunch) and of the drive bunch (magenta). The first column of Fig.~\ref{Fig:04}(a, e, i) corresponds to $w_{tl} = 0$, the second column (b, f, j) to $w_{tl}^{(1)} = 20\rm \mu m$, the third column of Fig.~\ref{Fig:04}(c, g, k) to $w_{tl}^{(2)} = 50\rm \mu m$, and the fourth column of Fig.~\ref{Fig:04}(d, h, l) to $w_{tl}^{(3)} = 100\rm \mu m$. In the first column, the TL is absent (Fig.~\ref{Fig:04}(a)); the positron witness bunch locates in the accelerating phase of the axial field $E_z$ (Fig.~\ref{Fig:04}(e)) and in the focusing phase of the radial field $W_r$ (Fig.~\ref{Fig:04}(i)) over its entire height. In the second column, the TL width is small (Fig.~\ref{Fig:04}(b)), and the positron witness bunch again locates entirely in the accelerating phase of the axial field $E_z$ (Fig.~\ref{Fig:04}(f)) and in the focusing phase of the radial field $W_r$ (Fig.~\ref{Fig:04}(j)) over its entire height. In the third column, the TL thickness is close to the skin depth of the tubular plasma~\footnote{The plasma skin depth is $\delta=c/\omega_p$, where $c$ is the speed of light in vacuum and $\omega_p=\left(4\pi n_{pe}e^2/m_e\right)^{1/2}$ is the plasma (Langmuir) frequency, with $n_{pe}$ the plasma density and $e$ and $m_e$ the electron charge and mass. For the tubular plasma, $\delta_{p(2)} = 42.961\rm \mu m$.} (Fig.~\ref{Fig:04}(c)), and although most of the positron witness bunch is accelerated by the axial field $E_z$, its upper part (close to the bunch radius) is decelerated by the negative field there (Fig.~\ref{Fig:04}(g)). This occurs because the period of the axial field $E_z$ increases as a result of the reduced tubular-plasma density in the transition layer. At the same time, the focusing of the witness bunch by the radial field $W_r$ is almost unchanged (Fig.~\ref{Fig:04}(k)). Finally, in the fourth column the TL width exceeds the skin depth of the tubular plasma (Fig.~\ref{Fig:04}(d)). The tubular-plasma density within the transition layer drops so strongly, increasing the period of the axial field $E_z$, that only the lower half of the positron witness bunch is accelerated while its upper part is decelerated (Fig.~\ref{Fig:04}(h)). Nevertheless, the focusing of the witness bunch by the radial field $W_r$ not only persists but even improves slightly (Fig.~\ref{Fig:04}(l)). As for the electron witness bunch, as the TL thickness changes it never leaves the region of negative axial field $E_z$ over its entire height and therefore continues to be accelerated (Fig.~\ref{Fig:04}(e), (f), (g) and (h)). The radial wakefield $W_r$ acting on the electron witness bunch remains almost unchanged as the transition-layer thickness varies and stays positive (focusing for electrons) across the bunch cross-section (i.e., for $0\leq r\leq R_b$) (Fig.~\ref{Fig:04}(i), (j), (k) and (l)).

\begin{figure}[!tbh]
  \centering
  \includegraphics[width=0.80\textwidth]{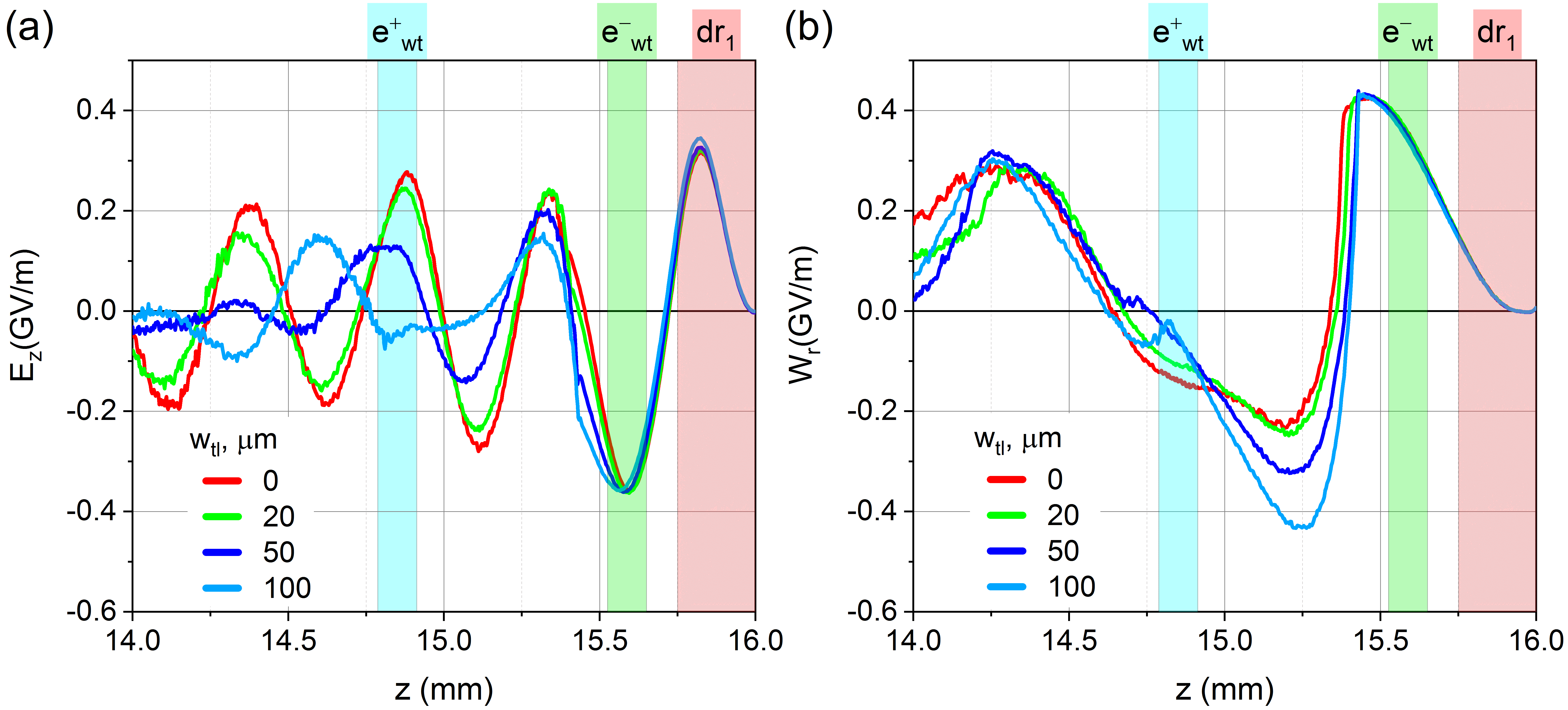}
  \caption{Longitudinal profiles of (a) the axial wakefield $E_z(r=R_b)$ and (b) the radial wakefield $W_r(r=R_b)$ excited by the driver, as functions of the longitudinal coordinate $z$ at the moment it reaches the waveguide's output, for different transition-layer thicknesses $w_{tl}$. The colored rectangles mark the positions of the witness bunches (cyan --- positron, green --- electron) and of the driver (pink).}\label{Fig:05}
\end{figure}
Figure~\ref{Fig:05} shows (a) the axial wakefield $E_z$ and (b) the radial wakefield $W_r$ excited by the drive bunch, as functions of the longitudinal coordinate $z$ at $r=R_b$, for different transition-layer thicknesses $w_{tl}$. Figure~\ref{Fig:05}(a) shows that an increase in the transition-layer thickness leads to a longer period of the accelerating axial field $E_z$ and to a change in its amplitude compared with the case in which the TL is absent. In the region of the positron witness bunch, the axial field $E_z$ for $w_{tl} \neq 0$ is smaller than in the case $w_{tl} = 0$, or even becomes negative (i.e., decelerating); in contrast, in the region of the electron witness bunch the amplitude of $E_z$ for $w_{tl} \neq 0$ is larger than for $w_{tl} = 0$, and the phase of $E_z$ locates close to the region of optimal acceleration for the electron bunch. As a result, as the transition-layer thickness increases, the acceleration of the positron witness bunch deteriorates while that of the electron witness bunch improves, as can be seen in Fig.~\ref{Fig:03}. As shown in Fig.~\ref{Fig:05}(b), the period of the radial field $W_r$ is almost unaffected by the increase in transition-layer thickness. The value of $W_r$ in the region of the electron witness bunch also remains unchanged. However, in the region of the positron witness bunch, a sharp peak appears on the curve $W_r(z)$ as the width of the transition region increases, leading to some reduction in the focusing of the positron bunch. This peak arises at the location where the flow of electrons from the region of tubular plasma within the transition layer passes by the positron witness bunch (see the region near the intersection of the horizontal dashed line and the cyan vertical dashed line for $e_{wt}^+$ on the color map of $n_{pe}/n_{p(1)}$ in Fig.~\ref{Fig:04}(d)).

\begin{figure}[!tbh]
  \centering
  \includegraphics[width=0.80\textwidth]{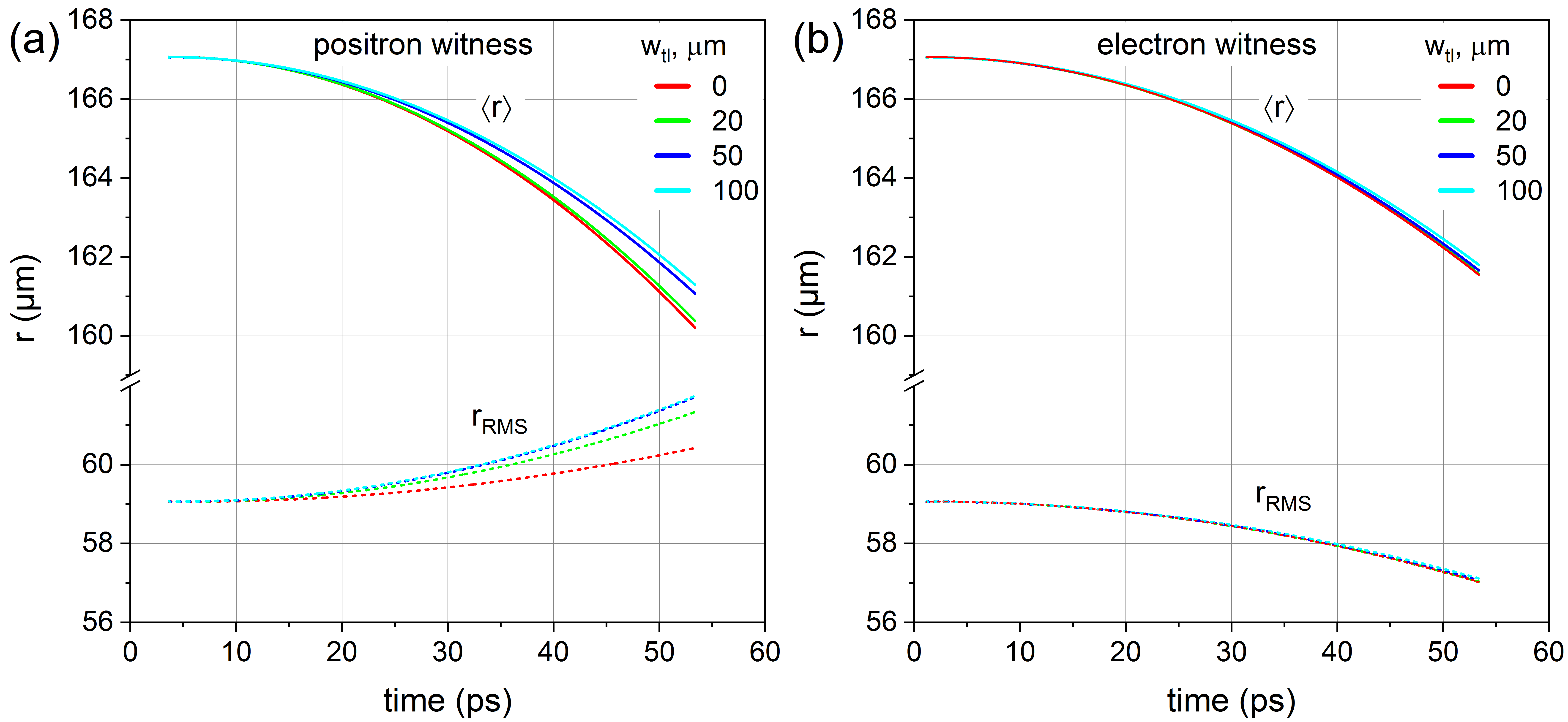}
  \caption{Time evolution of the mean radii $\langle r \rangle$ and of the root-mean-square deviations of the radii from their mean values, $r_{RMS}$, for the (a) positron and (b) electron witness bunches, for different transition-layer thicknesses $w_{tl}$.}\label{Fig:06}
\end{figure}
Figure~\ref{Fig:06} shows the time evolution of the mean radii $\langle r \rangle$\footnote{$\langle r \rangle = \sum\limits_i r_i\cdot q_i/Q$, where $r_i$ and $q_i$ are the radius and charge of a macroparticle and $Q$ is the total bunch charge.} and of the root-mean-square deviations of the radii from their mean values, $r_{RMS}$\footnote{$r_{RMS} = \sqrt{\sum\limits_i (r_i-\langle r \rangle )^2\cdot q_i/Q}$.}, for the (a) positron and (b) electron witness bunches for different transition-layer thicknesses. It can be seen that the decrease in the mean radius $\langle r\rangle$ of the positron test bunch is larger than that of the electron bunch, which is related to the stronger focusing of the positron witness bunch. The transverse spatial spread of the particles, $r_{RMS}$, also increases for the positron witness bunch (Fig.~\ref{Fig:06}(a)) but decreases for the electron witness bunch (Fig.~\ref{Fig:06}(b)) as they move along the accelerator, and this spread is larger for the positron bunch. This is related to the larger amplitude and stronger nonlinearity, along the transverse coordinate $r$, of the radial wakefield acting on the positron witness bunch compared with the radial field in which the electron witness bunch is located (see the third row of plots in Fig.~\ref{Fig:04}). The amplitude of the radial wakefield in the region of the positron witness bunch increases toward the axis, leading to a substantial increase in the on-axis positron density, which raises $r_{RMS}$ with time. For the field $W_r$ in the region of the electron witness bunch, an almost linear decrease is observed as the radius $r$ decreases. As a result, the change in the electron bunch density in the radial direction is small compared with the initial uniform value. The weak changes in $\langle r \rangle$ and $r_{RMS}$ for different transition-layer thicknesses $w_{tl}$ observed for the electron witness bunch (Fig.~\ref{Fig:06}(b)), compared with the changes in these quantities observed for the positron witness bunch (Fig.~\ref{Fig:06}(a)), are related to the small changes in the radial field for different values of $w_{tl}$ in the region occupied by the electron witness bunch (see Fig.~\ref{Fig:05}(b)). Nevertheless, despite these features, the presence of the transition layer does not cause defocusing of either the positron or the electron witness bunch.

As the results in Figs.~\ref{Fig:03} and~\ref{Fig:06} show, the presence of the transition layer strongly affects the acceleration of the positron bunch (especially for a thick transition layer) but has a weak effect on the acceleration and focusing of the electron witness bunch. Therefore, the optimization of parameters to improve the acceleration and transport of the witness bunch is carried out only for the positron bunch.

Thus, the main effect of the transition layer is to increase the period of the accelerating axial field, which in turn is related to the reduced tubular-plasma density within the transition layer. At the same time, the change in the plasma-column density within the transition layer has practically no effect on the period of the radial wakefield. However, the presence of the transition layer can cause irregular distortion of the radial wakefield due to the flow of electrons moving from the boundary region of the tubular plasma into the less dense on-axis plasma.

In order to optimally accelerate the positron bunch in the LPW with a transition layer by the wakefield of the electron bunch, while preventing an increase in the transverse size of the bunches, one can start from the analytically calculated parameters of the LPW without a transition layer and modify some of them in a specific way. Two possible options for such modifications in the presence of a transition layer are proposed: option (I) --- change (increase) the tubular-plasma density $n_{p(2)}$, or option (II) --- increase the witness-bunch injection delay time $t_{del}$ and change (decrease) the plasma-column density $n_{p(1)}$.

Let us consider these modifications in more detail. In option (I), for a given transition-layer thickness we increase the tubular-plasma density $n_{p(2)}$ in small steps relative to the value given in Table~\ref{tab:01}, computing at each step the witness-bunch energy gain $\langle\Delta E\rangle$ at the time $t = 53.37\,ns$, when the drive bunch reaches the output of the structure. It is assumed that increasing the plasma density $n_{p(2)}$ at each successive step increases the witness-bunch energy gain $\langle\Delta E\rangle$. If, instead, increasing the plasma density $n_{p(2)}$ decreases the energy gain $\langle\Delta E\rangle$, the increase in $n_{p(2)}$ is stopped, and a polynomial is fitted through the points $\left[n_{p(2)}^i,\,\langle\Delta E\rangle^i\right]$, where $i$ is an index, $n_{p(2)}^i$ is the abscissa, and $\langle\Delta E\rangle^i$ is the ordinate of each point. The maximum of the polynomial gives the largest witness-bunch energy gain $\langle\Delta E\rangle$, and its abscissa gives the optimal tubular-plasma density $n_{p(2)}^{tl}$, which provides the greatest witness-bunch acceleration for the given transition-layer thickness. In option (II), the increased delay time $t_{del}^{tl}$ for a given transition-layer thickness is found in the same way as in~\cite{Markov2022}. Thereafter, assuming that the time lag is proportional to the Langmuir frequency of the plasma column and knowing the squared ratio $\left(t_{del}^{e^+}/t_{del}^{tl}\right)^2$, where $t_{del}^{e^+}$ is the delay time given in Table~\ref{tab:01}, one can reduce the plasma-column density by the same factor, $n_{p(1)}^{tl} = n_{p(1)}\cdot\left(t_{del}^{e^+}/t_{del}^{tl}\right)^2$, to obtain optimal focusing of the witness bunch.

\subsection{Obtaining the optimal tubular-plasma density}
\begin{figure}[!tbh]
  \centering
  \includegraphics[width=0.9\textwidth]{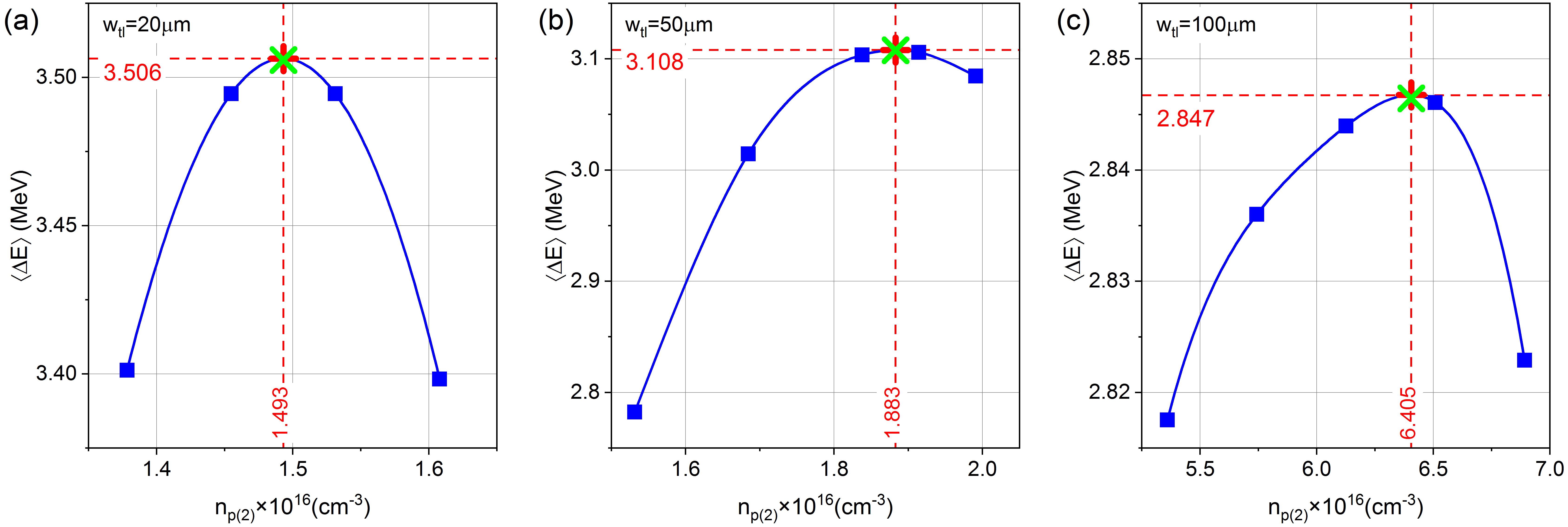}
  \caption{Mean energy gain of the positron witness bunch, reached at $t=53.37\,ns$, as a function of the tubular-plasma density for different transition-layer widths: (a) $w_{tl} = 20\,\mu m$, (b) $w_{tl} = 50\,\mu m$ and (c) $w_{tl} = 100\,\mu m$. The red dashed lines show the coordinates of the maximum of the polynomial function $\langle\Delta E\rangle\left(n_{p(2)}\right)$. The green cross marks the energy gain of the witness bunch achieved in the simulation of the accelerating structure with the optimal tubular-plasma density $n_{p(2)}^{tl}$.}\label{Fig:07}
\end{figure}
Figure~\ref{Fig:07} shows how the optimal tubular-plasma densities $n_{p(2)}^{tl}$ are found using the algorithm of option (I). The blue curves show the polynomial fits of the positron witness bunch energy gain $\langle\Delta E\rangle\left(n_{p(2)}\right)$ for several points with different tubular-plasma densities, in the region where the growth of the energy gain turns into a decrease. Table~\ref{tab:02} lists the resulting values of $n_{p(2)}^{tl}$.

\begin{table}[!tbh]
\caption{Optimal tubular-plasma density $n_{p(2)}^{tl^{\mathstrut}}$ for different transition-layer thicknesses $w_{tl_{\mathstrut}}$, obtained using the algorithm of option (I)}
\centering
\begin{tabular}{c | c}
\hline
  $w_{tl},\,\rm \mu m$ & $n_{p(2)_{\mathstrut}}^{tl}\times10^{16},\,\rm cm^{-3^{\mathstrut}}$\\
\hline
  20 & 1.49\\
  50 & 1.88\\
  100 & 6.41\\
\hline
\end{tabular}
\label{tab:02}
\end{table}

\begin{figure}[!tbh]
  \centering
  \includegraphics[width=0.40\textwidth]{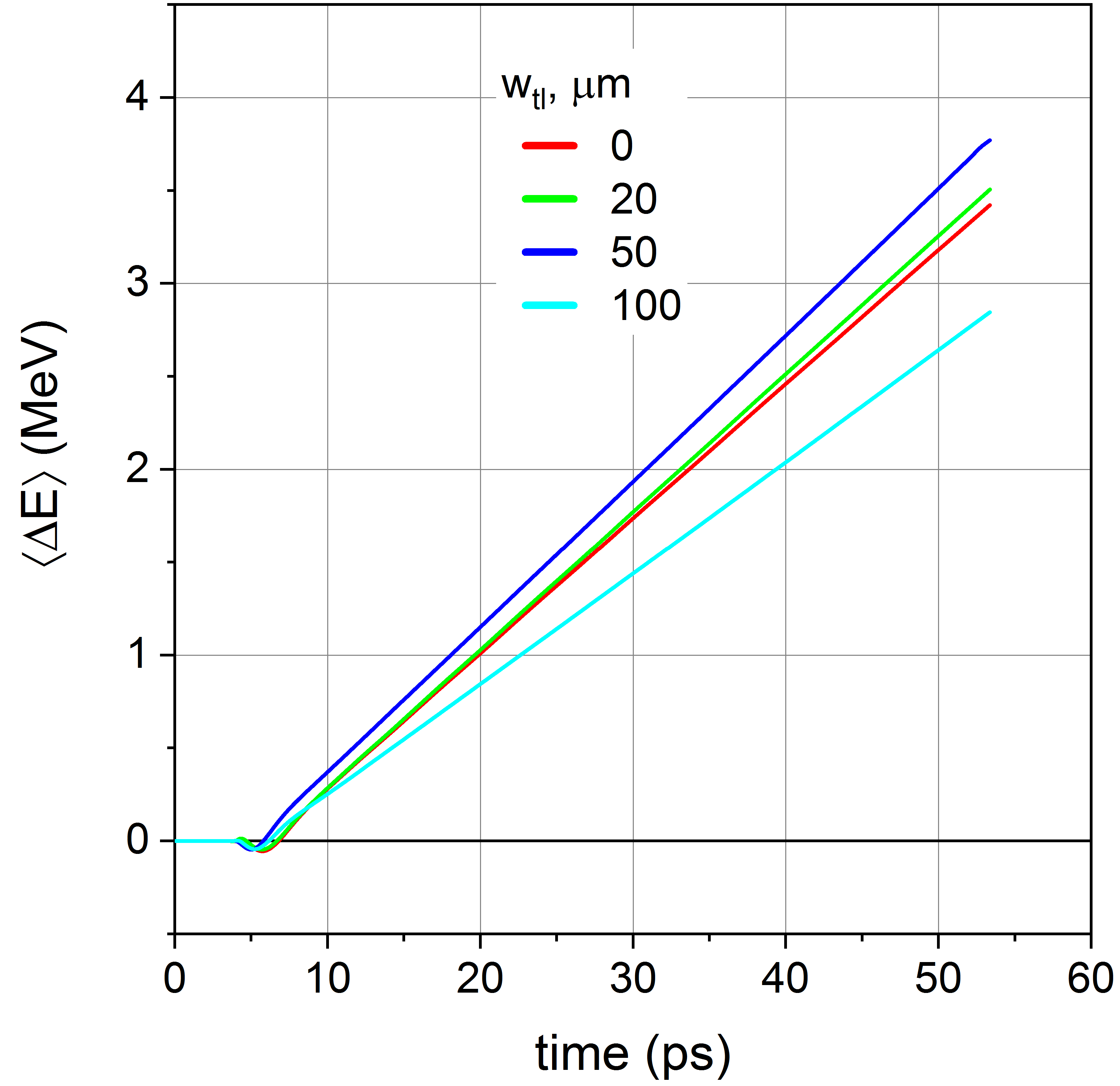}
  \caption{Change in the mean energy of the positron witness bunch as it moves in the wakefield generated by the electron driver, for different transition-layer widths and the optimal tubular-plasma density values $n_{p(2)}^{tl}$ given in Table~\ref{tab:02}.}\label{Fig:08}
\end{figure}
Figure~\ref{Fig:08} shows the change in the mean energy gain $\langle\Delta E\rangle$ as a function of time for the positron witness bunch as it moves in the wakefield generated by the electron driver, for different transition-layer widths and the values of $n_{p(2)}^{tl}$ given in Table~\ref{tab:02}. Comparing Figs.~\ref{Fig:03} and~\ref{Fig:08}, it can be seen that choosing the optimal tubular-plasma density makes it possible to increase the mean energy of the accelerated positron bunch.

\begin{figure}[!tbh]
  \centering
  \includegraphics[width=0.40\textwidth]{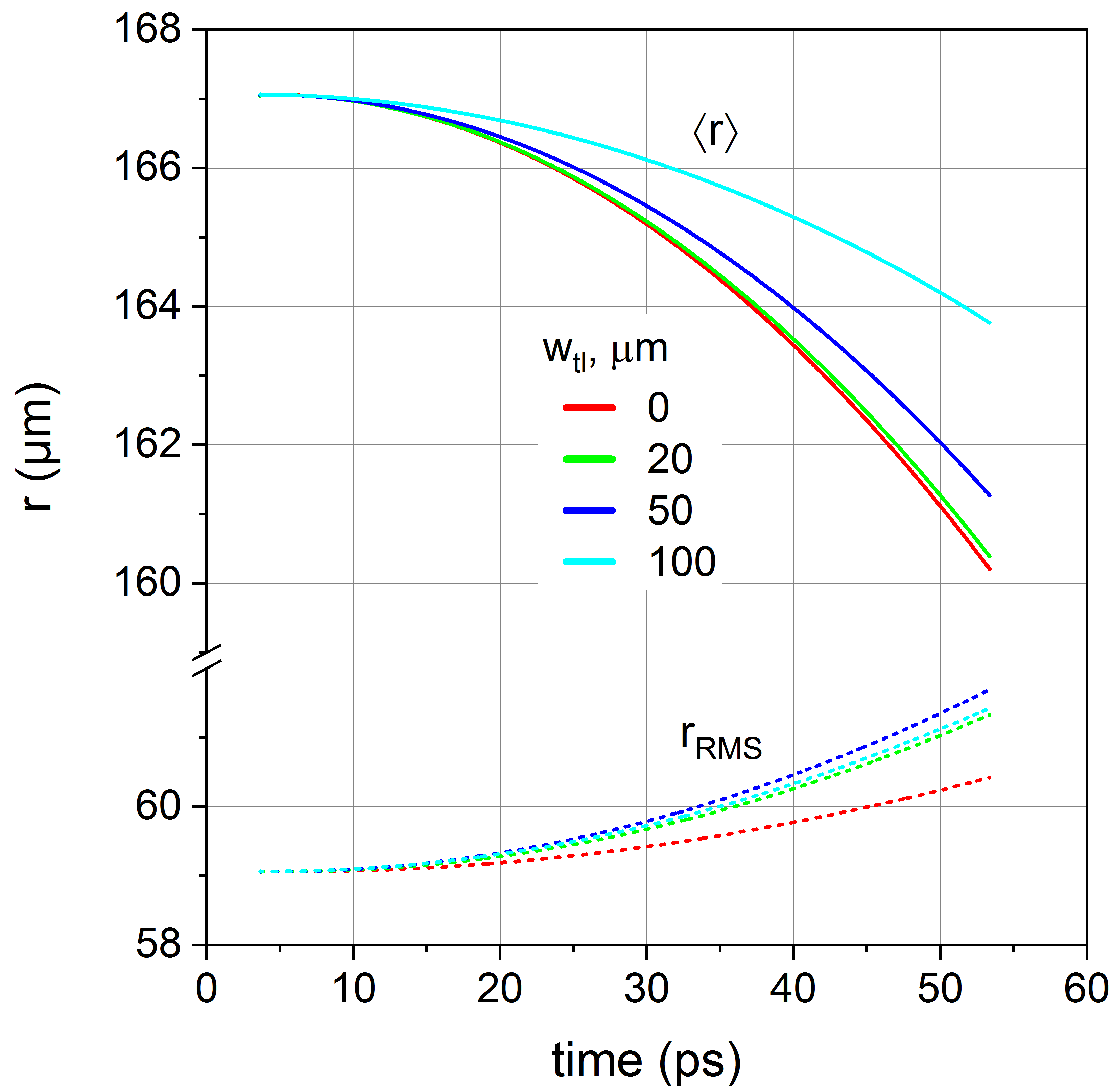}
  \caption{Time evolution of the mean radius $\langle r \rangle$ and of the root-mean-square deviation of the radius from its mean value, $r_{RMS}$, for the positron witness bunch, for different transition-layer thicknesses $w_{tl}$ and the optimal tubular-plasma density values $n_{p(2)}^{tl}$ given in Table~\ref{tab:02}.}\label{Fig:09}
\end{figure}
To verify the focusing condition for the accelerated bunch, $\langle r \rangle$ and $r_{RMS}$ were calculated for the positron witness bunch for different transition-layer thicknesses $w_{tl}$ and the optimal tubular-plasma density values $n_{p(2)}^{tl}$ given in Table~\ref{tab:02}. The result is shown in Fig.~\ref{Fig:09}. It can be seen that, at $w_{tl} = 100\,\rm \mu m$, the curve of $\langle r \rangle$ for the positrons in Fig.~\ref{Fig:09} deviates less from its initial value at $w_{tl} = 0$ than in Fig.~\ref{Fig:06}, because at the tubular-plasma density $n_{p(2)}^{tl} = 6.41\cdot10^{16}\rm cm^{-3}$ the radial wakefield in the region occupied by the positron witness bunch is substantially lower than at $n_{p(2)} = 1.53\cdot10^{16}\rm cm^{-3}$, resulting in weaker focusing of the bunch particles. Otherwise, the differences between the dependences shown in Fig.~\ref{Fig:09} and Fig.~\ref{Fig:06} are minor.

\subsection{Obtaining the optimal delay time and plasma-column density}
The method for finding the optimal delay time $t_{del}$ that maximizes the acceleration of the witness bunch in the wakefield generated by the driver is described in~\cite{Markov2022} and is, in principle, no different from the procedure for determining the optimal tubular-plasma density described above. The resulting optimal values of the delay time $t_{del}^{tl}$ and of the plasma-column density $n_{p(1)}^{tl}$ for the acceleration of the positron witness bunch, obtained using the algorithm of option (II), are given in Table~\ref{tab:03}.

\begin{table}[!tbh]
\caption{Optimal delay times $t_{del}^{tl}$ and plasma-column densities $n_{p(1)}^{tl}$ for accelerating the positron witness bunch at different transition-layer thicknesses $w_{tl}$, obtained using the algorithm of option (II)\\[2pt]}
\centering
\begin{tabular}{c | c | c}
\hline
  $w_{tl},\,\rm \mu m$ &  $t_{del}^{tl},\,\rm ps$ &$n_{p(1)_{\mathstrut}}^{tl^{\mathstrut}}\times10^{14},\,\rm cm^{-3}$\\
\hline
  20 &  3.679 & 4.80\\
  50 &  3.858 & 4.37\\
  100 & 4.352 & 3.43\\
\hline
\end{tabular}
\label{tab:03}
\end{table}

\begin{figure}[!tbh]
  \centering
  \includegraphics[width=0.40\textwidth]{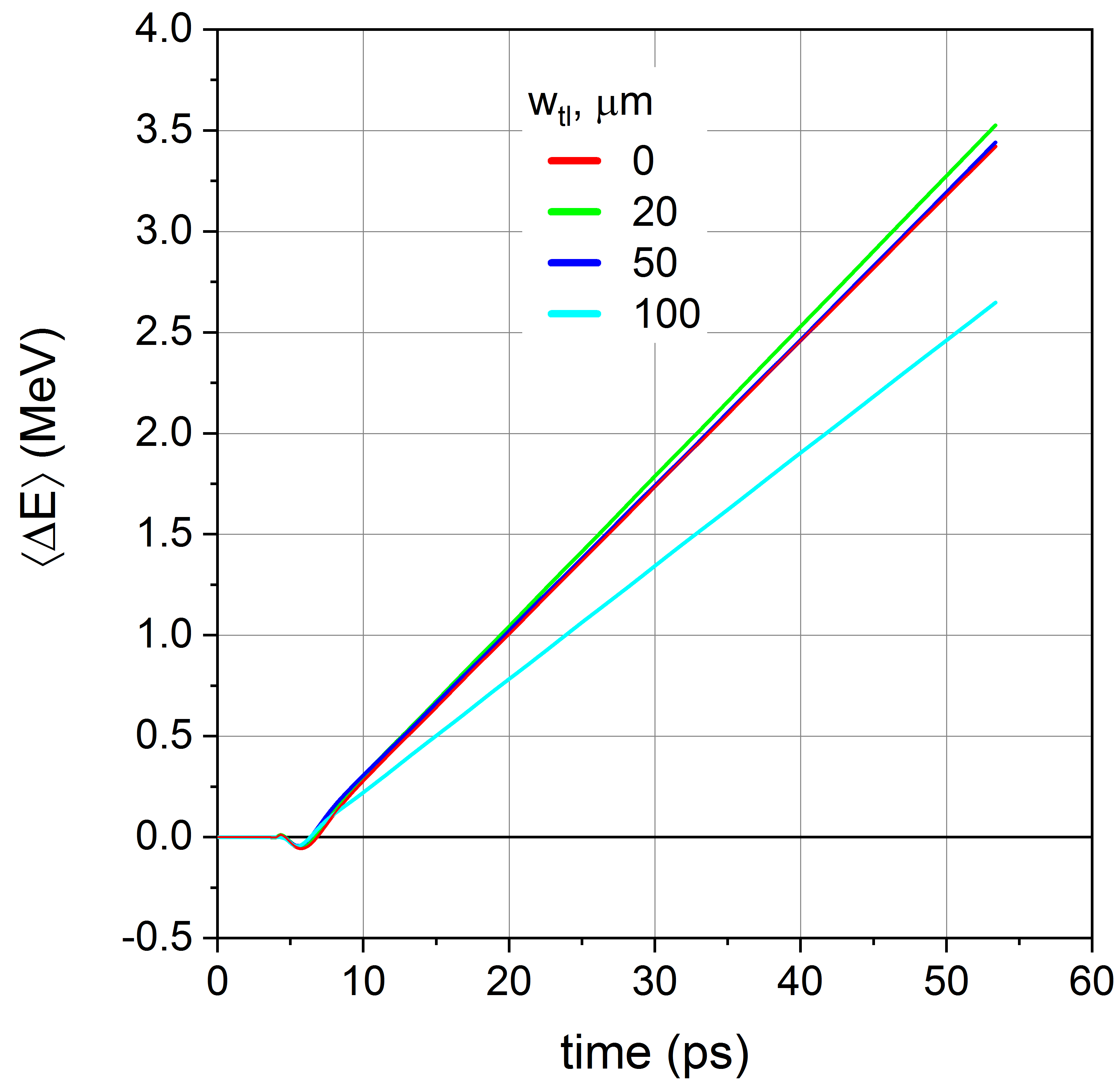}
  \caption{Change in the mean energy gain of the positron witness bunch as it moves in the wakefield generated by the electron driver, for different transition-layer widths and the optimal delay times $t_{del}^{tl}$ given in Table~\ref{tab:03}.}\label{Fig:10}
\end{figure}
Figure~\ref{Fig:10} shows the change in the mean energy gain $\langle\Delta E\rangle$ as a function of time for the positron witness bunch as it moves in the wakefield generated by the electron driver, for different transition-layer widths and the optimal delay times $t_{del}^{tl}$ given in Table~\ref{tab:03}. Comparing Fig.~\ref{Fig:03}(a) and Fig.~\ref{Fig:10}, it can be seen that choosing the optimal delay times increases the energy of the accelerated positron bunch.

\begin{figure}[!tbh]
  \centering
  \includegraphics[width=0.80\textwidth]{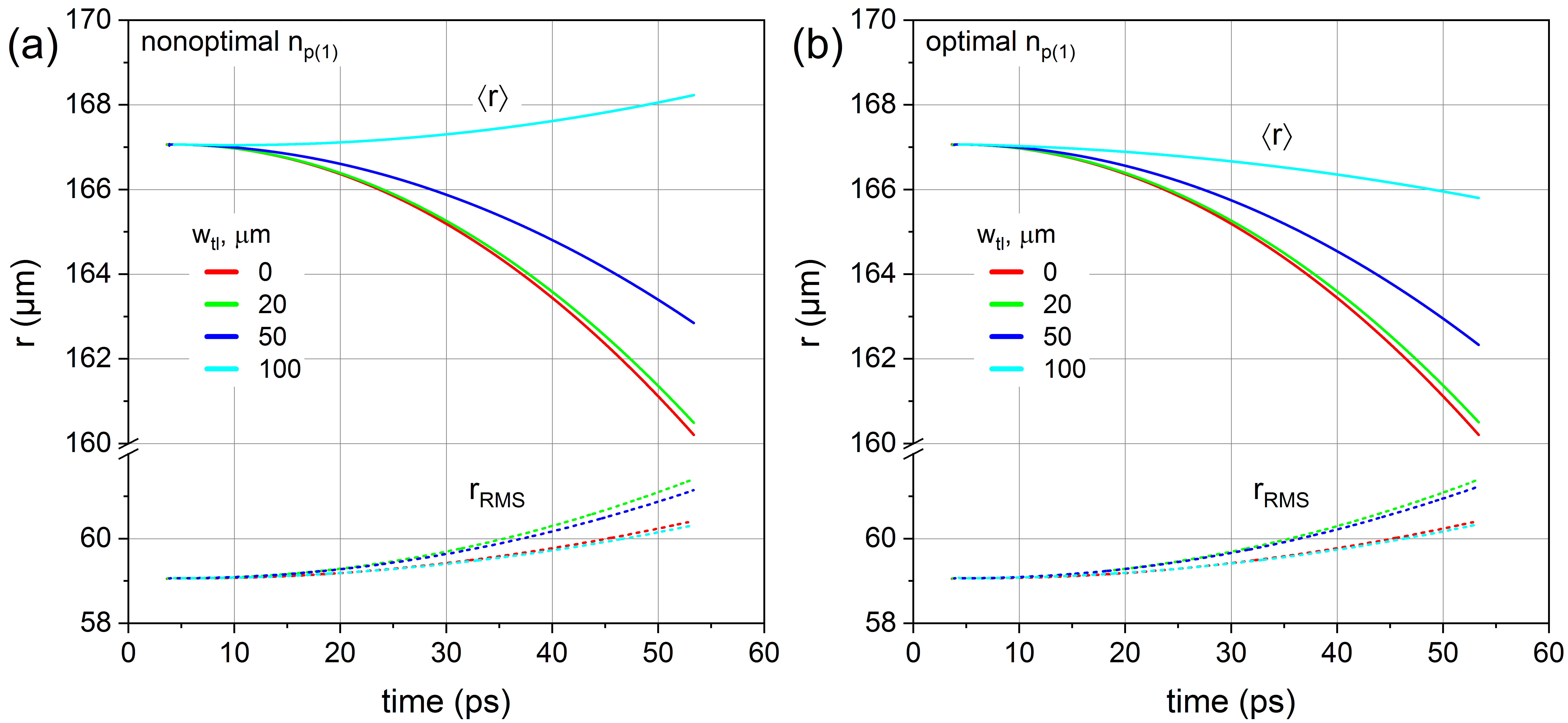}
  \caption{Time evolution of the mean radius $\langle r \rangle$ and of the root-mean-square deviation of the radius from its mean value, $r_{RMS}$, of the positron witness bunch for different transition-layer thicknesses $w_{tl}$, for (a) the non-optimal plasma-column density $n_{p(1)}=4.93\cdot 10^{14}\,\rm cm^{-3}$ and (b) the optimal density values $n_{p(1)}^{tl}$ given in Table~\ref{tab:03}.}\label{Fig:11}
\end{figure}
Figure~\ref{Fig:11} demonstrates the need to change the plasma-column density $n_{p(1)}$ for radially stable acceleration of the positron witness bunch in the wakefield structure formed by the layered plasma with a transition layer. Figure~\ref{Fig:11}(a) shows that for the non-optimal plasma-column density $n_{p(1)}=4.93\cdot 10^{14}\,\rm cm^{-3}$, calculated for the case of a layered plasma with a sharp boundary and no transition layer, the mean bunch radius $\langle r \rangle$ increases with time for the wide transition layer $w_{tl}=100\,\rm \mu m$, i.e., the transverse size of the positron witness bunch grows, whereas for the optimal plasma-column density values $n_{p(1)}^{tl}$ given in Table~\ref{tab:03}, shown in Fig.~\ref{Fig:11}(b), no such growth occurs and the bunch is focused as it is accelerated.

\section{Summary}~\label{sec:5}
This paper presents the results of numerical studies of the effect of a transition layer between two plasma layers of different densities, filling a cylindrical waveguide, on the electromagnetic field generated by a relativistic electron bunch. The structure of the wakefield, as well as the acceleration of the positron and electron bunches, was analyzed for different transition-layer widths. It is shown that the presence of a transition layer in the plasma filling under study reduces the amplitude and modifies the structure of the excited wakefield, resulting in a decrease in the energy gain of the accelerated bunches and an increase in their transverse size. The methods have been proposed and successfully demonstrated for adjusting the parameters of the accelerator structure and the bunches (specifically, the tubular plasma density, the plasma column density, and the injection time) to compensate for the adverse effect of the transition layer on the acceleration and focusing of test bunches.

\section*{Acknowledgments}
The study is supported by the National Research Foundation of Ukraine under the program “Excellent Science in Ukraine” (project No. 2023.03/0182).

\bibliographystyle{elsarticle-num}
\bibliography{BibliographyURL}
\end{document}